# Quantum and Classical Magnetoresistance in Ambipolar Topological Insulator Transistors with Gate-tunable Bulk and Surface Conduction

Jifa Tian[1,2,*], Cuizu Chang[3,4], Helin Cao[1,2], Ke He[3], Xucun Ma[3], Qikun Xue[4] and Yong P. Chen[1,2,5,#]

1. Department of Physics, Purdue University, West Lafayette, Indiana 47907, USA

2. Birck Nanotechnology Center, Purdue University, West Lafayette, Indiana 47907, USA

3. Beijing National Laboratory for Condensed Matter Physics, Institute of Physics, Chinese Academy of Sciences, Beijing 100190, P. R. China

4. State Key Laboratory for Low-Dimensional Quantum Physics, Department of Physics, Tsinghua University, Beijing 100084, P. R. China

5. School of Electrical and Computer Engineering, Purdue University, West Lafayette, Indiana 47907, USA

*Email: tian5@purdue.edu; #Email: yongchen@purdue.edu

**Abstract**

Weak antilocalization (WAL) and linear magnetoresistance (LMR) are two most commonly observed magnetoresistance (MR) phenomena in topological insulators (TIs) and often attributed to the Dirac topological surface states (TSS). However, ambiguities exist because these phenomena could also come from bulk states (often carrying significant conduction in many TIs) and are observable even in non-TI materials. Here, we demonstrate back-gated ambipolar TI field-effect transistors in $(Bi_{0.04}Sb_{0.96})_2Te_3$ thin films grown by molecular beam epitaxy on $SrTiO_3(111)$, exhibiting a large carrier density tunability (by nearly 2 orders of magnitude) and a metal-insulator transition in the bulk (allowing effectively switching off the bulk conduction). Tuning the Fermi level from bulk band to TSS strongly enhances both the WAL (increasing the number of quantum coherent channels from one to peak around two) and LMR (increasing its slope by up to 10 times). The SS-enhanced LMR is accompanied by a strongly nonlinear Hall effect, suggesting important roles of charge inhomogeneity (and a related classical LMR), although existing models of LMR cannot capture all aspects of our data. Our systematic gate and temperature dependent magnetotransport studies provide deeper insights into the nature of both MR phenomena and reveal differences between bulk and TSS transport in TI related materials.

Topological insulators (TIs) are an exotic state of quantum matter with nominally-insulating bulk and spin-momentum-locked Dirac fermion conducting surface states, promising potential applications in both nanoelectronics and spintronics.[1-3] Many interesting phonemona based on such topological surface states (TSS) have been proposed, such as Majorana fermions [4,5], exciton condensation [6], topological magnetoelectric effect [7], etc. $Bi_2Se_3$, $Bi_2Te_3$, and $Sb_2Te_3$ have been proposed and identified as prototype 3D TIs, possessing TSS with a linear dispersion of energy vs momentum in the bulk band gap (BBG).[8-11] However, most of these commonly studied TI materials inevitably have impurities or defects during their growth, resulting in a doped bulk which can account for a significant part of the conductance, making it difficult to study and utilize the novel electronic transport of the TSS. To access the surface transport properties of the 3D TIs, various strategies have been attempted to suppress the bulk conduction, for example by (compensation) doping, increasing the surface-to-volume ratio, or electric gating. A ternary compound $(Bi_{1-x}Sb_x)_2Te_3$, an alloyed mixture of $Bi_2Te_3$ and $Sb_2Te_3$, has been shown as a promising group of 3D TIs with excellent tunability of the electronic properties (from n-type to p-type) by varying the composition $x$ [12,13]. Recently, the quantum anomalous Hall effect [14] has been successfully observed in Cr doped $(Bi_xSb_{1-x})_2Te_3$, adding further interests to examine the underlying electronic transport properties of this ternary TI system.

Most transport studies of TIs have focused on magnetotransport [15], particularly magnetoresistance (MR). In a few experiments, Shubnikov–de Haas oscillations [11,16-18] with $\pi$ Berry phase were observed and used as direct transport evidence for TSS Dirac fermions. However, the majority of TIs (particularly alloyed mixtures) do not show such quantum oscillations because of disorder or inhomogeneity. Instead two other MR features have been commonly used in a large number of experiments to probe the transport signature of TSS: the weak antilocalization (WAL, enhanced conductance resulting from quantum interference between coherent time-reversed closed paths) at low magnetic ($B$) field [19-29] and linear MR (LMR) [30-34] at high $B$. However, given the parallel conducting surface and bulk channels often existing in TIs, ambiguities can arise when attributing WAL and LMR to TSS, as both phenomena could also arise from bulk states, and have even been observed in many non-TI-based spin-orbit-coupled or narrow-gap semiconductors [35,36]. For example, it is often difficult to exactly determine the roles of TSS in earlier WAL studies in TIs with metallic bulk conduction, where the reported WAL conductance correction from only one coherent channel likely reflects the strongly mixed contributions from bulk and surfaces coupled together [19,24-27]. More recent experiments utilizing gate tuning of the surface-bulk coupling and the number of phase coherent channels in $Bi_2Se_3$ thin films [21,22] have pointed out a possibility to extract the transport signatures of the TSS through careful analysis. Despite these progresses, most experiments so far still dealt with significant bulk conduction, and did not reach or clearly demonstrate the so-called "topological transport" regime, where bulk is truly insulating and surface dominates the conduction. In addition to the ambiguities associated with WAL, the nature of LMR [13,30-34,36-41] itself has also been debated (with both classical [38,39] and quantum models [40,41] proposed) since its observation in silver chalcogenides and other narrow gap semiconductors [36,37]. The recent observations of LMR in TIs [30-34] have been often interpreted in the framework of the quantum LMR [30,32] due to the gapless energy spectrum of TSS, although it is not fully clear if the quantum model [40,41] applies without a clean separation of bulk and TSS. A better understanding of the nature of LMR will benefit its many proposed applications in areas such as spintronics or magnetoelectric sensors [32].

In this work, we perform a systematic transport study of ternary $(Bi_{0.04}Sb_{0.96})_2Te_3$ TI thin films (10 nm) grown by molecular beam epitaxy (MBE) on $SrTiO_3$ (STO)(111) substrates (250µm, Figs. 1a,b). Using STO (with its very high dielectric constant $\varepsilon_r$ at low temperature) as a back gate, we demonstrate a large tunability of the carrier density (by nearly 2 orders of magnitude) and Fermi level in our TI film, exhibiting an ambipolar field-effect (FE). This allows us to realize a gate-tuned metal-to-insulator transition in the bulk of our TI film, thus tuning the transport from the bulk-dominated regime (where the Fermi level is in the bulk valence band, BVB) to the topological transport regime (where Fermi level is in the BBG and TSS) with surface-dominated conduction at low temperature. We also systematically map out the gate and temperature dependent WAL and LMR (along with Hall measurements), revealing differences between the bulk and topological surface transport regimes and providing more insights on the natures of such MR phenomena.

The high quality $(Bi_{0.04}Sb_{0.96})_2Te_3$ (10 nm-thick) thin films studied here are grown by MBE on heat-treated 250µm-thick insulating STO (111) substrates [12]. The schematic of the sample is shown in Fig. 1a. Previous ARPES measurements have demonstrated that the TSS exists in the BBG and the $E_f$ is located below the Dirac point, indicating a p-type doping in the as-grown films [12]. The representative device structures are defined by standard e-beam lithography (EBL), followed by dry etching using Ar plasma. The Hall bar electrodes of the devices are fabricated by another EBL process followed by e-beam deposition of Cr/Au (5/80 nm). A Cr/Au (5/100 nm) film is e-beam deposited on the back of the STO substrate working as a back gate. Transport properties are measured by the conventional four-probe lock-in technique with an AC driving current of 100 nA at 17.77 Hz. In a typical device (shown in Fig 1b), the driving current is applied between electrodes "1,4" and the longitudinal resistance $R_{xx}$ and Hall resistance $R_{xy}$ are measured between electrodes "5,6" and "3,5", respectively. All the measurements are carried out in a variable temperature ($T$, from 1.4 K to 230 K) cryostat with a magnetic field $B$ (perpendicular to the film) up to ±6 T.

The temperature dependence of the zero $B$ field longitudinal resistance ($R_{xx}$ vs $T$) measured at different back gate voltages $V_g$ (from -60 V to 130 V) are presented in Fig. 1c. The STO substrate, with its very high relative dielectric constant $\varepsilon_r$ at low $T$ [13, 42], gives a strong gate modulation to the sample's carrier density. By increasing $V_g$, the Fermi level ($E_f$) can be tuned from the BVB to BBG (intercepting TSS, insets of Fig. 1c). For $V_g$ = -60 V, where $E_f$ is located in the BVB, the corresponding $R_{xx}$ (solid black curve) decreases with decreasing $T$ and saturates at low $T$, demonstrating a characteristic metallic bulk conduction. The temperature below which $R_{xx}$ appears to saturate moves to a smaller value (dashed yellow curve) at $V_g$ =-10 V, suggesting weakened metallic behavior. Further increasing $V_g$ to -5 V, $R_{xx}$ shows a clear upturn below ~30 K, indicating the appearance of an insulting behavior (attributed to freezing-out of thermally excited bulk carriers) in the film. Such a bulk insulating behavior can be significantly enhanced by further lifting $E_f$ into the BBG (green dotted line, $V_g$ =0 V) and eventually, for $V_g$ > 10 V the bulk insulating behavior onsets at a $T$ as high as ~100 K, with $R_{xx}$ approximately saturating with very weak $T$-dependence for $T$ <30 K. The bulk insulating behavior observed in $V_g$=0 V is consistent with the ARPES observation of $E_f$ located in the BBG in as-grown films [12]. The nearly saturated $R_{xx}$ (terminating the insulating behavior) for $V_g$>10 V indicates a remnant conduction that can be attributed to the TSS in the BBG [11,16-18,43,44] dominating the charge transport at low $T$ (see also Fig. S1). Our observations demonstrate a striking transition from the metallic to insulating behavior in the bulk of such films, driven by $E_f$ (tuned by the back gate). This transition can also be regarded as that from a "topological metal" to a "topological insulator", and is foundational to our study to clarify the relative roles played by the bulk and TSS in MR features.

Fig. 1d shows the FE behavior ($R_{xx}$ vs $V_g$) measured at $B$=0 T and $T$=1.4 K. All the curves of $R_{xx}$ vs $V_g$ show ambipolar FE. For example, $R_{xx}$ of curve "1" (black) is weakly modulated by the back gate as $V_g$<-20 V, but increases significantly and reaches a peak of ~12 kΩ when $V_g$ is increased to ~$12$ V (the charge neutral point (CNP), $V_{CNP}$, showing a FE on-off ratio of 600%) before decreasing again upon further increasing $V_g$. Furthermore, an appreciable hysteresis in $R_{xx}$ vs $V_g$ depending on the $V_g$ sweeping history and direction is observed. This hysteresis is common for STO due to its nonlinear dielectric response close to ferroelectricity [42] and may also relate to the interface charge traps (defects) between the substrate and TI film. Among the data plotted, curve "1" (where $V_{CNP}$ ~12 V) represents the first $V_g$ sweep from -60 V to 60 V after initial cooling down to 1.4 K. The corresponding $R_{xx}$ in curve "1" is consistent with the $R_{xx}$ values (shown as boxes with crosses in Fig. 1d) at each $V_g$ (-60 V, -10 V, -5 V, 0 V, and 10 V) extracted from Fig. 1c at 1.4 K. However, $V_{CNP}$ is shifted to 50 V (curve "2", green) and 43 V (curve "3", yellow) as $V_g$ sweeps backward (from 60 V to -60 V) and forward (-60 V to 60 V) again, respectively. Repeating the $V_g$ sweeps from 60 V to -60 V (or -60 V to 60 V), the FE curve will stabilize and follow the FE curve "2" (or "3"), respectively. All the data presented later are taken after this stabilization and in a forward sweeping direction starting from -60 V to minimize this hysteresis effect (also because of this hysteresis, one should not directly compare the $V_g$'s in the following data with those in Fig.1c, instead $V_g$-$V_{CNP}$ or low-$T$ resistance values are better indicators of the sample state). The ambipolar FE in Fig. 1d suggests a sign change of dominant charge carrier from p-type to n-type as $V_g$ crosses $V_{CNP}$, confirmed by the corresponding gate dependent Hall resistance $R_{xy}$ (exhibiting a sign change) and longitudinal resistance $R_{xx}$ measured at $B$=-6 T as shown in Fig. 1e. We note that the charge carriers are holes at $V_g = 0$ V, also consistent with the ARPES-measured $E_f$ position in as-grown films [12].

Figure 2 shows the gate-dependent $\Delta R_{xx}$ and the corresponding Hall resistance $R_{xy}$ (plotted as functions of magnetic field, $B$) at various temperatures (see also Fig. S2 for representative color plots showing simultaneously the gate and temperature dependence). Here, we define $\Delta R_{xx}(B)=R_{xx}(B)-R_{xx}(B=0\ T)$. In Fig. 2(a), all curves of $\Delta R_{xx}(B)$ obtained at different $V_g$ (from -60V to 60V) at $T$=1.4 K show a gate-dependent cusp at $|B|$ <1.5 T, a clear signature of the WAL. The amplitude of the cusp can be significantly enhanced by varying $V_g$ from -60V to 60V to tune $E_f$ from BVB to TSS, and reaches a maximum at CNP. Another interesting observation is the largely LMR observed at higher $B$ field in $\Delta R_{xx}(B)$, which is also strongly enhanced by gating towards TSS and will be discussed in more detail later. Meanwhile, the corresponding $R_{xy}$ vs $B$ also shows a strong gate dependence, as seen in the lower panel of Fig. 2a, with two main observations with increasing $V_g$: 1) the slope of $R_{xy}$ vs $B$ initially increases (related to the reduced carrier density) and is followed by a drop as well as a sign change while $V_g$ crosses the CNP, a direct manifestation of the sign change of charge carriers (from holes to electrons); 2) the corresponding shape of $R_{xy}$ vs $B$ changes from linear to non-linear, suggesting a change from one-band to two (or multiple) band transport (due to, for example, coexisting surface and bulk channels of opposite carriers, and/or electron and hole puddles). The $\Delta R_{xx}$ and $R_{xy}$ also show significant temperature dependences as shown in Figs. 2b-f. As $T$ increases from 1.4 K, the WAL cusp gradually weakens (Fig. 2b upper panel) and finally disappears at ~25 K, where the LMR becomes prominent and starts from very low B (<~ 0.2 T) for most of $V_g$'s (Fig. 2c upper panel and a zoomed-in view in Fig. S3). Further increasing $T$ (>~40K), $\Delta R_{xx}(B)$ becomes parabolic at low magnetic field ($B$<2 T), which becomes increasingly evident at further elevated $T$ as shown in Figs. 2d-f, with LMR still clearly observable at higher $B$ (>2 T). The ambipolar (sign change of slope, thus of charge carriers) and nonlinearity (near CNP) behaviors observed in $R_{xy}$ also become increasingly evident with increasing $T$ up to ~25 K (Figs. 2a-c). However, for $T$ >40 K, the observed $R_{xy}$ vs $B$ is always linear and has no sign change (Figs. 2d-f), indicating one-band behavior with p-type carriers. We note that the gate becomes less effective as the dielectric constant $\varepsilon_r$ of STO substrate becomes significantly reduced [42,45] at elevated $T$ (see also Fig. S4), rendering the ambipolar field effect no longer achievable. Our results map out a systematic evolution of both $R_{xx}(B, V_g, T)$ and $R_{xy}(B, V_g, T)$, demonstrating the transport properties in such system are highly dependent on $E_f$ (modulated by gating) and the temperature.

Further studies of the field and Hall effects as well as carrier density and mobility are presented in Fig. 3. Figs. 3a,b show the temperature dependences of $R_{xx}$ vs $V_g$ (measured at B=0 T) and $R_{xy}$ vs $V_g$ (measured at $B$=-6 T), respectively. Consistent with Fig. 2, for $T$ up to ~ 25 K, we see again the ambipolar FE (Fig. 3a) while $V_{CNP}$ increases from 45 V (at 1.4 K) to 80V (at 25 K), related to the decreased $\varepsilon_r$ of STO substrate mentioned above [42]. The corresponding $R_{xy}$ (Fig. 3b) also demonstrates the ambipolar behavior up to ~25 K, where $R_{xy}$ (initially negative) decreases with increasing $V_g$, followed by an upturn and sign change as $V_g$ crosses CNP. For $T > 40$ K, both $R_{xx}$ and $|R_{xy}|$ monotonously and weakly increase as $V_g$ increases with no indication of ambipolar behavior. We extract the carrier density ($n$) and Hall mobility ($\mu$) from $R_{xy}$ and $R_{xx}$ (in Fig. 2) using the one-band model at different $V_g$ and $T$ in the regime of p-type carriers (mostly form BVB, see below) where a linear $R_{xy}$ vs $B$ is observed, and plot the results in Figs. 3c,d. The carrier (holes) density ($n_p$) is ~1.8x10$^{14}$ cm$^{-2}$ at $V_g$ =-60 V & $T$=1.4 K. As $V_g$ increases and approaches CNP, $n_p$ decreases approximately linearly. The similar trend is also observed at higher $T$'s up to 25 K, while the slope of $n_p$ vs $V_g$ decreases significantly with $T$>25 K. An effective capacitance ($C$) per unit area of STO can be calculated from the slope and $C$ decreases from ~290 nF/cm$^2$ (corresponding to $\varepsilon_r$ ~82000) at 1.4 K down to ~5.3 nF/cm$^2$ ($\varepsilon_r$ ~1500) at 200 K (see Fig. S4), consistent with previously observed strongly $T$-dependent dielectric behavior of STO.[42,45] We also note that $n_p$ at $V_g$=0 V (Fig. 3c) decreases with increasing $T$. This is attributed to thermal excitation of $n$-type carriers, and confirms that the increased $V_{CNP}$ at higher $T$ (in Fig. 3a) is mainly due to the decreased STO capacitance. The temperature and gate-dependent mobility $\mu$ is shown in Fig. 3d. The mobility at 1.4 K increases with increasing $V_g$ (or decreasing $n_p$) from ~50 cm$^2$/Vs at $V_g$ =-60 V to ~140 cm$^2$/Vs at $V_g$ = 30 V. The similar behavior is observed up to 40 K, while for T>70 K, $\mu$ becomes ~110 cm$^2$/Vs and largely $V_g$ independent. The inset of Fig. 3d shows a summary of $\mu$ (in log scale) vs $n_p$ for all measured $T$'s, where the data appear to collapse together and can be fitted to $\mu \sim \mu_0 e^{-n_p/n_0}$ with $n_0 = 6 \times 10^{15}$ cm$^{-2}$, $\mu_0$=133.4 cm$^2$/Vs, suggesting that $\mu$ is mainly controlled by carrier density but not temperature (up to ~200 K) in our system. The measured density-dependent mobility may provide valuable input for understanding carrier transport and scattering mechanism in TI materials [46-49], important for developing TI-based devices. In the case where $R_{xy}$ is nonlinear with $B$ (seen in Fig. 2 and for many other $V_g$ and $T$'s not included in Fig. 3c) due to multiple conduction channels and coexisting holes and electrons, the one-band model will not yield accurate carrier density (as seen in Fig. S5, *n* calculated from such one-band fits starts to deviate from linear $V_g$-dependence close to CNP). While a multiple-band model [11,17,43,50] can in principle be applied to fit the non-linear $R_{xy}$, we found however, such an analysis does not give unique fitting results (yielding significant uncertainties) in our case. Applying one-band model fitting (Fig. S5) for our data measured at T=1.4 K, the lowest carrier (electron) density |*n*| achieved in our sample is ~$5.5 \times 10^{12}$ cm$^{-2}$ (an overestimate for the actual density) at $V_g$=60 V (where $R_{xy}$ is n-type and only slightly non-linear). This value is smaller than the estimated maximum electron density (>$1.2 \times 10^{13}$ cm$^{-2}$, see Fig. S5 caption for more details) that can be accommodated in the TSS before populating BCB, demonstrating that $E_f$ is already located in the upper part (above Dirac point, DP) of the TSS at $V_g$=60 V, *T*=1.4 K. This also demonstrates that we can reach a regime where the charge carriers are mostly from the TSS and band bending near the surface is not significant to populate the bulk carriers (otherwise such a low carrier density will not be reached). The corresponding *n* vs $V_g$ at low *T* (Fig. S5) demonstrates that we have successfully tuned $E_f$ from the BVB, to the lower and the upper parts of TSS (through DP), as the gate voltage is increased from -60 V to 60 V.

Now we present the gate and temperature effects on the WAL, which is a manifestation of quantum coherent transport in the low-*B* MR and observed in our sample below 15 K. Figure 4a shows the sheet conductance correction $\Delta G_\Box(B)=G_\Box(B)-G_\Box(B=0T)$ vs *B* at various gate voltages measured at 1.4 K, where $G_\Box=(L/W)/R_{xx}$ with *W* and *L* being the width and length of the channel (between voltage probes) respectively. While both the bulk and TSS of TI possess strong spin-orbit coupling and can give rise to WAL, we have observed that WAL is significantly enhanced as $E_f$ is tuned into the BBG to suppress the bulk conduction and decouple the top and bottom surfaces. The Hikami-Larkin-Nagaoka (HLN) [51] equation (1) has been widely applied to analyze $\Delta G_\Box(B)$ due to WAL:

$$\Delta G_\Box(B) \simeq \alpha \cdot \frac{e^2}{\pi h}\left[\psi\left(\frac{1}{2}+\frac{B_\phi}{B}\right) - \ln\left(\frac{B_\phi}{B}\right)\right] \quad (1),$$

where $\alpha$ is a prefactor expected to be -1/2 for a single coherent channel, $\psi$ is the digamma function, and $B_\phi = \hbar/4eL_\phi^2$ is a characteristic field (with $L_\phi = \sqrt{D\tau_\phi}$, the phase coherence length, and $\tau_\phi$ phase-coherence time, *D* the diffusion constant). It is found [22] that even with parallel conducting channels (such as bulk and surfaces), the total $\Delta G_\Box$ may still be fitted using Eq. (1) in terms of an effective $\alpha$ that reflects the inter-channel coupling (with A=2|$\alpha$| representing the effective number of coherent channels). Our data in Fig. 4a agrees well (up to 2 T) with the HLN fittings (dashed lines) (see also Fig. S6). The extracted values of |$\alpha$| ($\alpha$<0) and $L_\phi$ at various gate voltages and temperatures are shown in Figs. 4b-c, respectively. We find that |$\alpha$| is strongly gate-tunable and exhibits an "ambipolar" behavior (peaks ~ 1 at CNP) for all *T*'s where WAL is observed, revealing three regimes of behavior as the number of coherent channels and degree of inter-channel coupling are tuned by the gate: (I) $V_g$<-10 V, where $E_f$ is located in the BVB, $|\alpha| \sim 0.5$ indicates the surface and bulk are fully coupled into one coherent conduction channel (A~1); (II) As $V_g$ is increased (-10 V<$V_g$<+45 V) to lift $E_f$ toward the BBG (TSS), $|\alpha|$ (and A) increases, indicating the top and bottom surfaces start to decouple from the bulk and each other, toward forming two channels; when $V_g$ reaches ~45 V, where the $E_f$ is located close to CNP, $|\alpha|$ reaches a maximum ~1=|-(1/2+1/2)|) with A~2, corresponding to two fully separated phase coherent channels (surrounding the bottom and top surfaces); (III) Further increasing $V_g$> 45 V to increase the $E_f$ above the CNP in TSS and towards conduction band, $|\alpha|$ starts to decrease from 1, indicating the top and bottom surfaces start to be coupled again (via the bulk). Our analysis also suggests that care must be taken when attributing WAL to TSS (especially for $\alpha \sim -0.5$) [19,24-27] in a TI material with both bulk and surface conduction. In addition, we find that the phase coherent length $L_\phi$ also shows interesting gate dependence (Fig. 4c). In regime I, in contrast to the weak gate dependence of $|\alpha|$, $L_\phi$ at low $T$ (1.4 and 2.8 K) notably increases with increasing $V_g$ and peaks at $V_g$=-10 V. Between regimes II and III, $L_\phi$ reaches a local minimum when $V_g$ is near CNP (40 V-45 V), where $|\alpha|$ reaches a peak.

Figs. 4b-c also demonstrate the temperature dependence of $|\alpha|$ and $L_\phi$ in different regimes of gate voltage ($E_f$). We see that in regimes II and III, $|\alpha|$ is relatively insensitive to $T$. In regime I, $|\alpha|$ moderately decreases below 0.5 with increasing $T$ when the TSS is coupled to the bulk at $V_g$=-60 V and -10 V, similar to the behavior previously found in samples with bulk-dominated conduction and explained as the suppression of WAL at high $T$ [22]. Fig. 4d shows the temperature dependences of $L_\phi$ at 4 representative gate voltages. Previous studies have commonly fitted $L_\phi(T)$ to a power law, with the expectation that electron-electron scattering would give $L_\phi$ proportional to $T^{-0.5}$ in a 2D system and $T^{-0.75}$ in a 3D system.[44,52,53] The dashed lines in Fig. 4d show such power law fittings giving $L_\phi$ proportional to $T^{-0.38}$, $T^{-0.57}$, $T^{-0.34}$, and $T^{-0.34}$ for $V_g$=-60V, -10V, +45V and +60V. We note that the fitted power-law $T$-dependences of $L_\phi$ in previous experiments range from $T^{-0.24}$ [54], $T^{-0.5}$ [44,55] to $T^{-0.75}$ [22] for different TI thin films, and ~$T^{-0.37}$ [56,57] or $T^{-0.5}$ [58,59] for TI nanowires, suggesting that the observed power law can depend on detailed material or electronic properties, possibly related to other dephasing processes (not just electron-electron scattering, thus not universal), therefore a gate-dependence as we see may not be unexpected. In our case, only at $V_g$=-10 V (where $L_\phi$ is also the largest, Fig. 4c) we observe $L_\phi \sim T^{-0.5}$, close to the predicted behavior of electron dephasing due to electron-electron scattering in 2D. However, strong deviation from this behavior is observed for other $V_g$'s, where $L_\phi$ is also shorter (for $V_g$=-60 V, $L_\phi \sim T^{-0.38}$, where only one coupled 2D conduction channel exists; for $V_g$= 45 V, where there are 2 decoupled conduction channels, and 60V, both giving $L_\phi \sim T^{-0.34}$), suggesting existence of additional dephasing

processes at these $V_g$'s (for example, such processes may be related to other carrier pockets in the valence band for $V_g$=-60 V; and electron-hole puddles near CNP for $V_g$= 45 and 60 V).

We now discuss the pronounced LMR observed. In contrast to the standard quadratic MR [41], which displays a relatively small magnitude and saturates at moderate magnetic fields, the observed LMR does not seem to saturate in high fields. Figs. 5a-d show the temperature-dependent LMR and corresponding $R_{xy}$ at two representative $V_g$'s. The LMR (measured at 6 T) in terms of relative MR ($\Delta R_{xx}(B)/R_{xx}(B=0T)$) varies from a few percent to ~30% depending on the gate voltage and $T$. We have noted the high-$B$ (>2T) MR to be slightly sub-linear (super-linear) for $T$<25 K ($T$>25 K) with the 25 K MR being closest to strictly linear (Figs. 5a,c as well as Fig. 2). Figure 5e shows the slope ($k$) of LMR (extracted between 3 T and 6 T) vs $V_g$ at different temperatures. When $E_f$ is in BVB ($V_g$<0 V), $k$ has very little $T$-dependence and weakly increases with increasing $V_g$, whereas $k$ is dramatically enhanced (by as much as 10 times, and becomes much more $T$-dependent) and approaches a maximum near CNP as $E_f$ is tuned into BBG (TSS) at low $T$. Such an observation is confirmed by the temperature dependence of $k$ at five $V_g$'s ($V_g$=-60, -10, 27.5, 45 and 60 V) plotted in Fig. 5f. For $V_g$= -60 V and -10 V, where $E_f$'s are located in the BVB, the slopes of LMR have little temperature dependence in the measured $T$ range. As $E_f$ is tuned into the BBG (TSS), the slope $k$ dramatically increases with the decreasing $T$ and reaches the highest value observed near CNP at $T$=1.4 K. Interestingly, as plotted in the inset of Fig. 5f, we find that $k$ vs $n_p$ (in log-log scale, only including data points with one-band $n_p$ for holes as those included in Fig. 3c) at different temperatures follow a similar trend with $k$ approximately proportional to $n_p^{-1}$ (except the highest $T$ data at 200 K), suggesting that the carrier density is important to control the slope of the LMR, which is significantly enhanced as $E_f$ approaches TSS (see also Fig. S8). We note that, at a fixed density $n_p$, $k$ shows little $T$ dependence up to 25 K, while $k$ notably decreases with increasing $T$ for $T$>25 K (Fig. S8). In the n-type regime (close to CNP), the large LMR is accompanied by prominent nonlinearity in $R_{xy}$ (Fig. 5d as well as Fig. 2c). This observation suggests that charge inhomogeneity may play an important role in the enhanced LMR, as discussed further below.

Our gate tunable WAL can also be understood [22] in terms of a competition between the phase coherence time ($\tau_\phi$, which does not vary strongly with the $V_g$, Fig. S7) and the surface-to-bulk scattering time [22] ($\tau_{SB}$, which decreases with increasing carrier density [60]), where the effective |$\alpha$| generally increases with increasing $\tau_{SB}/\tau_\phi$ [22] as $V_g$ is tuned towards CNP. When the $E_f$ is located in the BVB, $\tau_\phi$ (~ hundreds of ps, Fig. S7) is much larger than $\tau_{SB}$ (<< 1ps [60]), resulting in a single phase coherent channel. As $E_f$ is tuned into the TSS, $\tau_{SB}$ significantly increases due to the reduced carrier density and bulk conduction, and ultimately can become larger than $\tau_\phi$, realizing two decoupled channels. The very weak increase of |$\alpha$| at higher $T$ in regimes II and III seen in the inset of Fig. 4d may be attributed to a decrease

in $\tau_\phi/\tau_{SB}$, which increases the inter-channel decoupling, given that both $L_\phi$ and $\tau_\phi \sim L_\phi^2$ decreases as $T$ increases (Fig. S7) while the $\tau_{SB}$ should be relatively constant as both $R$ and the carrier density (Figs. 1c & 3c) change little up to ~15 K. However, in regime I, $|\alpha|$ moderately decreases below 0.5 at higher $T$, where $\tau_{SB}$ is expected to be much shorter than $\tau_\phi$ in the measured temperature range [22]. Such a decrease of $|\alpha|$ below 0.5 in presence of strong bulk conduction has been attributed to WAL getting suppressed when $\tau_\phi$ decreases and becomes comparable to the spin-orbit scattering time $\tau_{SO}$ at higher $T$ [22].

There have been two common models proposed for the LMR, the classical model by Parish-Littlewood (P-L) [38,39] (in terms of inhomogeneities in disordered conductors) and the quantum model by Abrikosov [40,41]. According to the quantum model [40,41], a LMR would occur at the quantum limit where the applied magnetic field is so large that only one [40,41] or few [37] Landau levels (LLs) are populated. This condition is more easily satisfied in a gapless semiconductor with linear energy-momentum dispersion [40,41]. Furthermore, the theory also predicts that $\Delta R_{xx}$ (magnitude of LMR) is proportional to $1/n^2$ ($n$ being the carrier density) and has no direct dependence on $T$ (as long as $T$ remains lower than the energy gap between LLs and the $E_f$). More recently, another model by Wang & Lei, based on the TSS and assuming uniform carrier density, relaxes the requirement of extreme quantum limit (instead it assumes many LLs are filled and smeared by disorder) and predicts a LMR with $\Delta R_{xx} \propto 1/n$ [61]. On the other hand, without invoking the linear dispersion spectrum, Parish and Littlewood proposed a classical mechanism for the LMR, as a consequence of potential and mobility fluctuations in an inhomogeneous electronic system, resulting in admixture of Hall resistance into $R_{xx}$ on a microscopic level and a LMR [38,39]. The classical model predicts that the relative MR = $\Delta R_{xx}/R_{xx}(B=0T)$ (thus slope $k$) should be proportional to a mobility scale $\mu_S=\max(|\mu|,|\Delta\mu|)$, where $\mu$ is mobility (with positive/negative sign for holes/electrons) and $\Delta\mu$ is the mobility fluctuation, and the cross over field $B_C$ (the magnetic field at which the MR curve changes from parabolic to linear) is proportional to $1/\mu_S$. The previously reported LMR in TIs have been often interpreted in terms of the quantum model [40,41] based on the linear dispersion of the TSS [30,32]. However, the studies reported so far have not systematically measured the dependence of LMR on carrier density, mobility and temperature, while such information is important to unambiguously identify and distinguish different mechanisms for LMR as discussed above. It also remains unclear whether bulk and surface carriers may contribute differently to LMR. We find that none of the existing models can fully explain our observed LMR. For example, in the p-type one-band carrier regime (inset of Fig. 5f), where $E_f$ is in the BVB (because the lowest $n_p$ ($1.7\times10^{13}$ cm$^{-2}$) extracted here is higher than the estimated maximum hole density ($1.2\times10^{13}$ cm$^{-2}$) that can be accommodated in the TSS before populating BVB), we are far from the extreme quantum limit (with LMR observable at very high LL filling factor, eg. > 8,000) assumed in Abrikosov's quantum model, and the assumptions of linear band dispersion or TSS as invoked by Abrikosov [41] or used in Wang-Lei model also do not apply. We have plotted the observed LMR amplitude (represented by $\delta R_{xx}=R_{xx}(6T)-R_{xx}(3T)$, focusing on the $B>3T$ regime where LMR is fully developed) vs $n_p$ (Fig. S9) and found it cannot be fitted to a single power law (either $1/n_p^2$ (Abrikosov) or $1/n_p$ (Wang-Lei)) over this density range, but rather appears to cross over from a $\sim 1/n_p$ behavior for $n_p>4\times10^{13}$ cm$^{-2}$ to $\sim 1/n_p^2$ behavior for $n_p<4\times10^{13}$ cm$^{-2}$ (except for the highest-$T$ data at 200K). As $E_f$ is tuned into the TSS or CNP, the LMR is enhanced and shows strong $T$-dependence (see Figs. 5e,f and also S9a,b), while concurrently $R_{xy}(B)$ becomes strongly non-linear (Fig. 5d, also Fig. 2c) and exhibits a sign-change (carrier type inversion), indicating charge inhomogeneity (such as coexisting electron and hole puddles) is significant in this ambipolar regime. This is at odds with the $T$-independent LMR predicted in both Abrikosov's quantum model and the Wang-Lei model (which also assumes uniformly distributed charge carriers), but instead suggests that charge inhomegeity (as highlighted in the classical mechanism) may play important roles in the LMR. To address the question whether the classical PL model can describe our observed LMR (in both BVB and TSS regimes), it is important to examine the correlation between the LMR slope $k$, cross over field $B_C$, and mobility (Figs. S10-S12). In the BVB regime (where p-type carrier $\mu$ can be extracted from one-band model), we find that $k$ appears to be approximately proportional to $\mu$ (consistent with PL model prediction if the mobility scale $\mu_s \sim \mu$) up to $\mu \sim 100$ cm$^2$/Vs, but becomes poorly correlated with $\mu$ for higher $\mu$ (Fig. S10). Fig. S11 shows the crossover field $B_C$ as a function of $\mu$, which is *qualitatively* ($B_C$ generally lower for larger $\mu$) but not quantitatively consistent with the PL model (which would predict $1/B_C$ to be proportional to $\mu$, if $\mu_s \sim \mu$). Furthermore, we note that PL model should predict $1/B_C$ to be proportional to $k$ (even without direct knowledge of $\mu_s$, which could depend on $\Delta\mu$). We have examined the correlation between $1/B_C$ and $k$ (Fig. S12) and find that while such a proportionality may hold approximately at relatively high $T$ (>40K), it does not hold for the full data set (including the 25K data, where LMR is particularly pronounced). In any case, our systematic data on LMR have revealed the following important points: 1) TSS can strongly enhance the LMR; 2) the charge inhomogeneity also plays important roles in the observed LMR, whose behaviors appear to be *qualitatively* captured by the classical P-L model but several aspects are still *not quantitatively* accounted for. A more complete model likely needs to take into account both the full band structure (bulk and TSS) and inhomogeneity in order to fully explain the observed LMR. Our systematic results on the density, mobility, and temperature dependences of LMR (Fig. 5, Fig. S8-12) can provide important insights for understanding the mechanisms of LMR and key inputs to develop a more complete model.

To conclude, we have systematically studied the transport, particularly MR as well as Hall effect of MBE grown TI (Bi$_{0.04}$Sb$_{0.96}$)$_2$Te$_3$ thin films on STO(111) substrates at different temperatures and gate voltages.

We demonstrate an ambipolar field effect showing a large tunability of the carrier density (by nearly 2 orders of magnitude), and we realize the topological transport regime with insulating bulk and surface-dominated conduction. Furthermore, as the Fermi level is driven into the bandgap where bulk conduction is suppressed and the system reaches charge neutral point, the amplitudes of both WAL (with the number of coherently coupled channels changing from 1 to 2) and LMR (also accompanied by strongly nonlinear Hall effect) are significantly increased. The data suggest that not only TSS strongly enhances LMR, but the charge inhomogeneity also plays important roles (as pointed out previously in a classical model for LMR). Our results map out a rich evolution of transport behavior from a doped topological metal to an intrinsic TI in such films, revealing important transport signatures of TSS and differences of the transport of topological insulators between the bulk conduction and topological surface transport regimes, as well as providing insights for the mechanism of LMR.

Acknowledgements

The work at Purdue is supported by DARPA MESO program (Grant N66001-11-1-4107). The MBE synthesis of TI thin films is supported by NSF of China (Grants 11134008 and 11174343).

**Figure Captions:**

**Figure 1 $(Bi_{0.04}Sb_{0.96})_2Te_3$ field effect device and its temperature and gate dependent transport**. (a) Schematic of MBE-grown $(Bi_{0.04}Sb_{0.96})_2Te_3$ thin films (thickness=10 nm) on $SrTiO_3$ (STO, thickness=250 μm, used as back gate); (b) Optical image of a fabricated Hall bar shaped device; (c) Temperature-dependent $R_{xx}$ (left axis, with corresponding $R_{xx}$ per square, $R_□$, plotted on the right axis) curves at various different $V_g$. All the curves are measured at zero magnetic field ($B$) and during the cooling process. The upper and lower insets show the schematic band structure with different Fermi level ($E_f$) positions for $V_g$=

130 V and -60 V, respectively; (d) $R_{xx}$ (left axis, with corresponding $R_\square$, plotted on the right axis) as a function of gate voltage ($V_g$) for different $V_g$ sweeping histories (all measured at $B$=0 T). The arrows label the $V_g$ sweeping directions. The symbols (boxes with crosses inside) label the resistances extracted from (c) for each $V_g$ at 1.4 K; (e) $R_{xx}$ and $R_{xy}$ as functions of $V_g$ at $B$ = -6 T and $T$=1.4 K. The unshaded (shaded) areas mark the $V_g$ ranges with $p$ ($n$) type dominant carriers.

**Figure 2 Magnetoresistance ($\Delta R_{xx}$=$R_{xx}(B)$-$R_{xx}(B$=$0T)$) and Hall resistance $R_{xy}$ vs $B$ for different $V_g$ at various temperatures**. (a) $T$=1.4 K; (b) $T$=15 K; (c) $T$=25 K; (d) $T$=40 K; (e) $T$=70 K; (f) $T$=200 K.

**Figure 3 Field effect and gate-dependent carrier density ($n_p$) and mobility ($\mu$) at various temperatures**. Temperature dependences of (a) longitudinal resistance $R_{xx}$ vs $V_g$ measured at $B$=0 T and (b) Hall resistance $R_{xy}$ vs $V_g$ measured at $B$=-6 T; (c) The carrier (hole) density $n_p$ and; (d) mobility $\mu$ extracted using a 1-band model from Fig. 2 at gate voltages where a linear Hall effect is observed. Inset is $\mu$ (in log scale) vs $n_p$ at different temperatures. Dashed line is an exponential fit ($\mu = \mu_0 e^{-n_p/n_0}$) to all data points with $n_0 = 6 \times 10^{15}$ cm$^{-2}$, $\mu_0$=133.4 cm$^2$/Vs.

**Figure 4 Gate and temperature dependent weak antilocalization (WAL) effect**. (a) Gate dependence of sheet conductance ($\Delta G_\square$=$G_\square(B)$-$G_\square(B$=$0T)$) at 1.4 K for various $V_g$'s. The dashed curves are fits to the data using the HLN model (Eq. 1); Gate dependent (b) prefactor $|\alpha|$=- $\alpha$ (A=2$|\alpha|$, right axis, is the number of coherent conducting channels) and (c) phase coherence length $L_\phi$ extracted from Eq. (1) at various temperatures; (d) Temperature dependence of $L_\phi$ and (inset) $|\alpha|$ measured at four representative $V_g$=60 V, 45 V, -10 V and -60 V. Dashed lines in the main panel (for $L_\phi$) are power law fittings (see text).

**Figure 5 Gate and temperature tunable linear magnetoresistance (LMR)**. The LMR, $\Delta R_{xx}/R_{xx}(B$=$0T)$, and the corresponding $R_{xy}$ as functions of magnetic field $B$ at (a,b) $V_g$ = -60 V and (c,d) $V_g$ = 60 V measured at various temperatures (ranging from 1.4 K to 200 K), respectively; (e) The gate voltage dependence of the extracted LMR slope ($k$, extracted between 3 T and 6 T) at different temperatures; (f) The LMR slope ($k$) vs $T$ for different gate voltages. Inset shows $k$ vs $n_p$ (in log-log scale) at different temperatures for $n_p$ values shown in Fig. 3c (one band p-type carriers). Gray band indicates a power law with exponent -1 ($k \sim n_p^{-1}$). The data for $T$=200 K can be fitted by $\sim n_p^{-2}$.

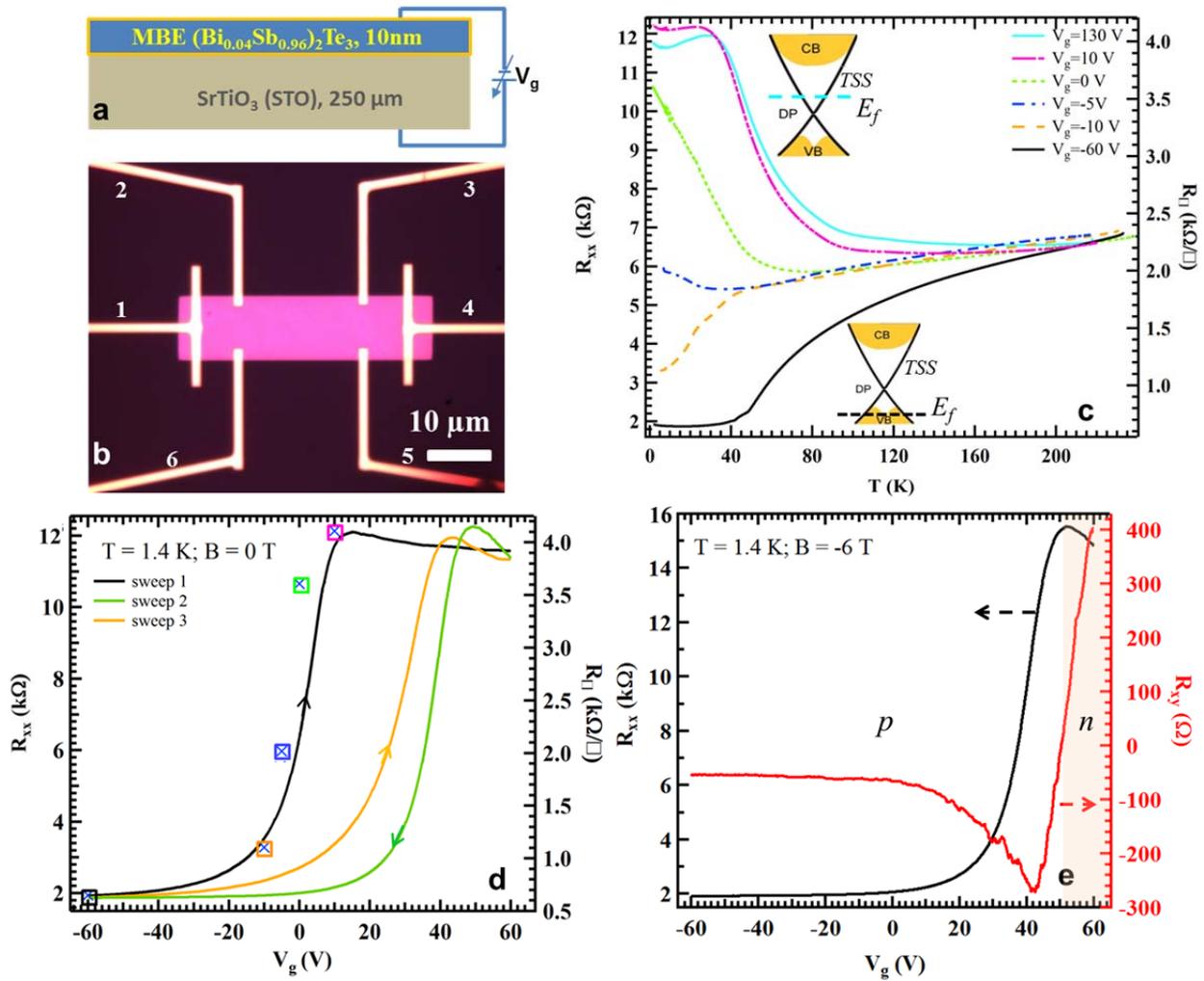

(Figure 1 Tian *et al*.)

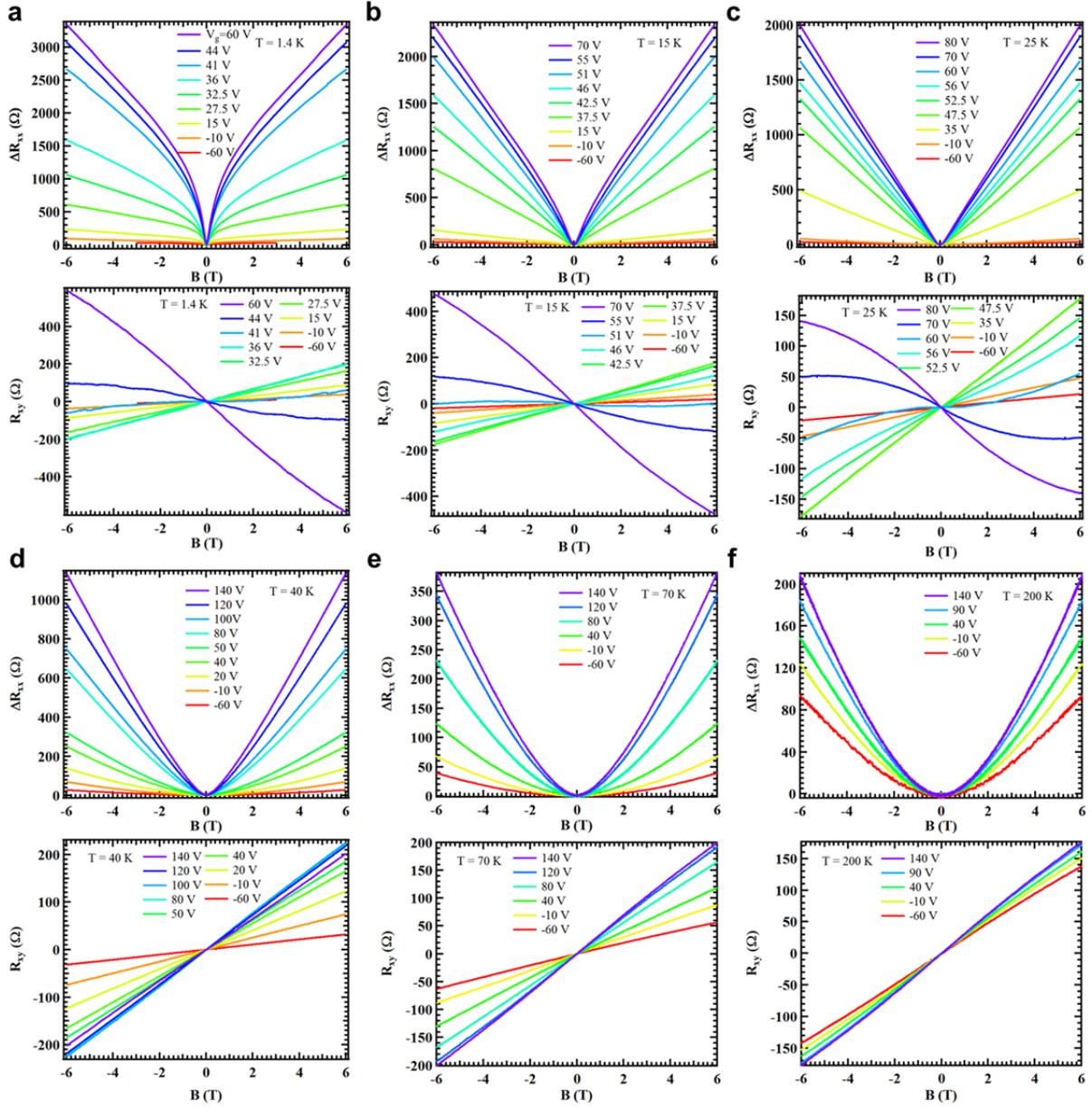

(Figure 2 Tian *et al*.)

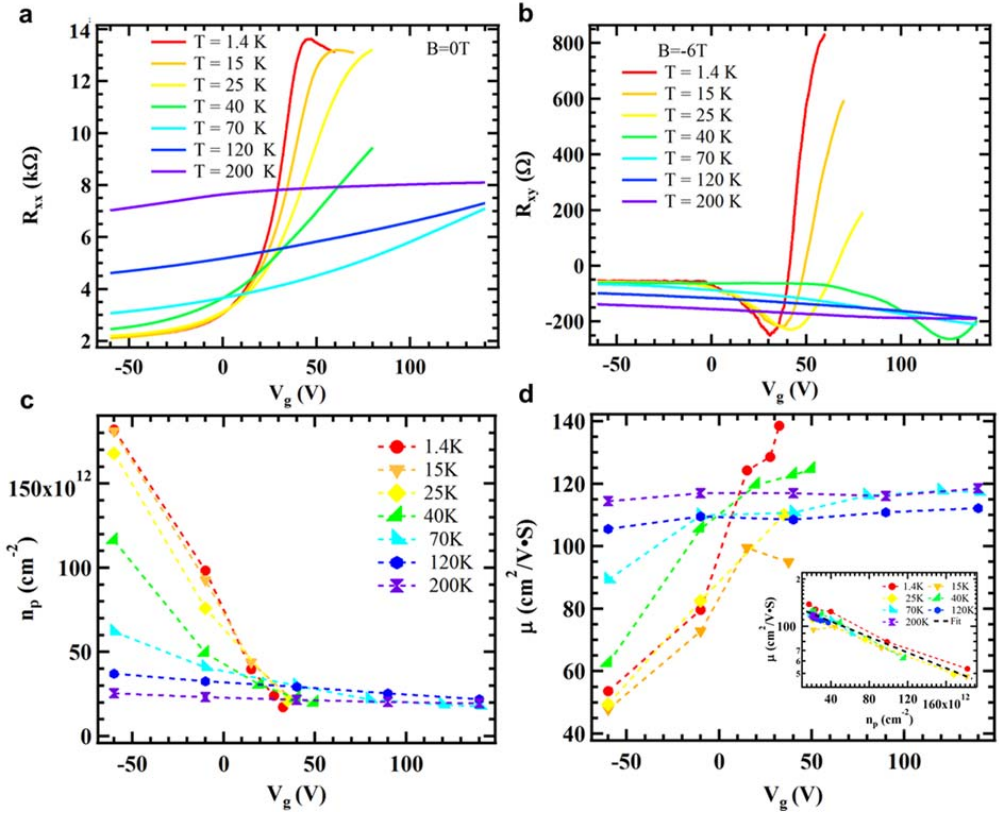

(Figure 3 Tian *et al*.)

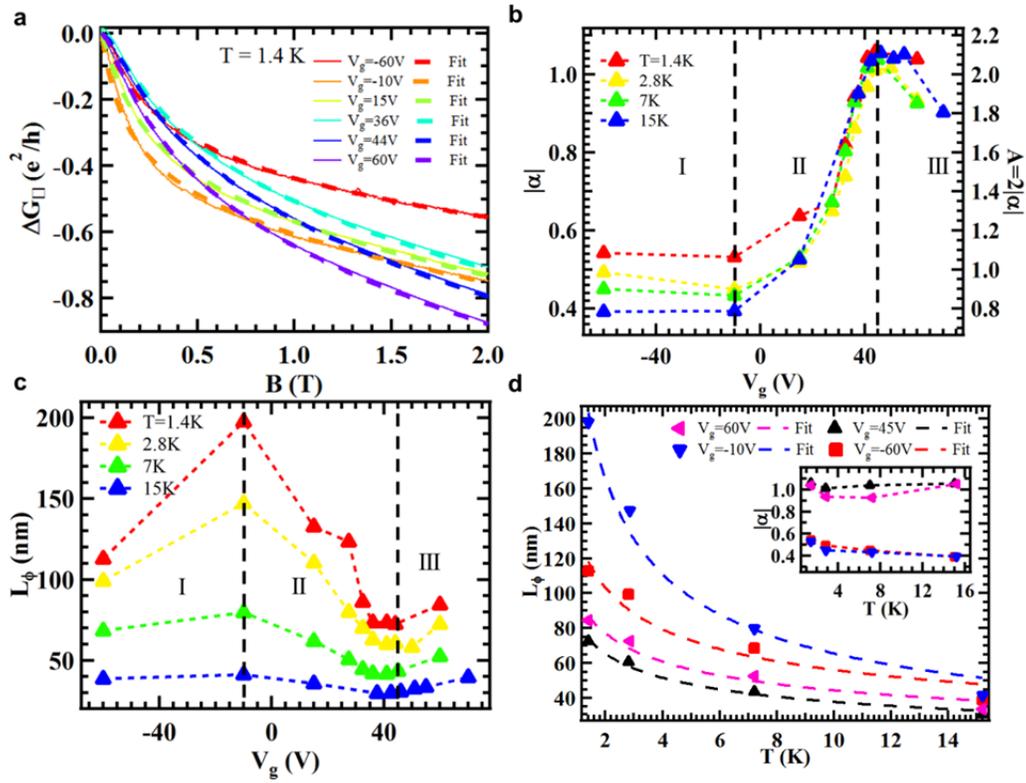

(Figure 4 Tian *et al.*)

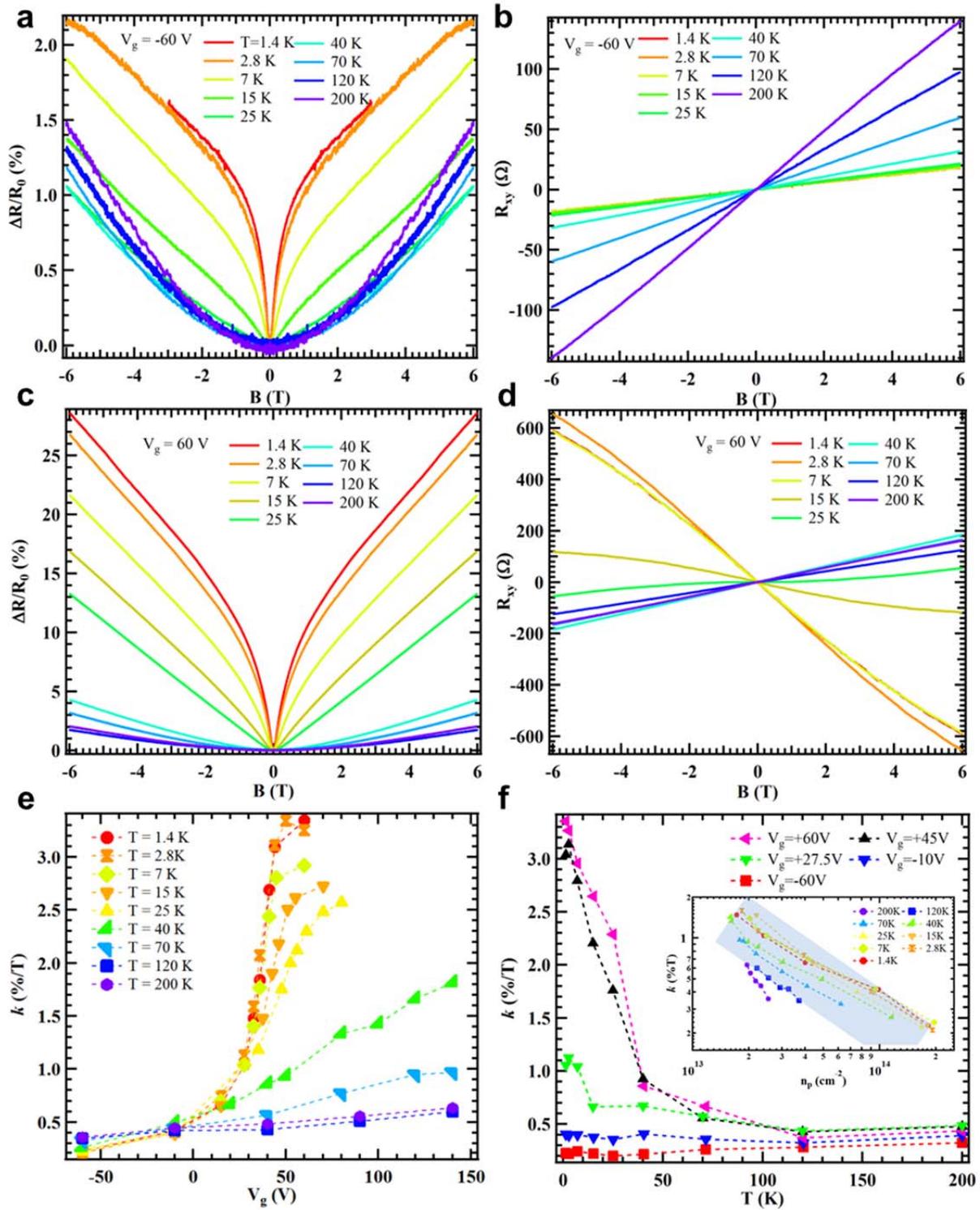

(Figure 5 Tian *et al.*)

Supplemental Materials

**Quantum and Classical Magnetoresistance in Ambipolar Topological Insulator Transistors with Gate-tunable Bulk and Surface Conduction**

Jifa Tian[1,2,*], Cuizu Chang[3,4], Helin Cao[1,2], Ke He[3], Xucun Ma[3], Qikun Xue[4] and Yong P. Chen[1,2,5,#]

1. Department of Physics, Purdue University, West Lafayette, Indiana 47907, USA

2. Birck Nanotechnology Center, Purdue University, West Lafayette, Indiana 47907, USA

3. Beijing National Laboratory for Condensed Matter Physics, Institute of Physics, Chinese Academy of Sciences, Beijing 100190, P. R. China

4. State Key Laboratory for Low-Dimensional Quantum Physics, Department of Physics, Tsinghua University, Beijing 100084, P. R. China

5. School of Electrical and Computer Engineering, Purdue University, West Lafayette, Indiana 47907, USA

*Email: tian5@purdue.edu; #Email: yongchen@purdue.edu

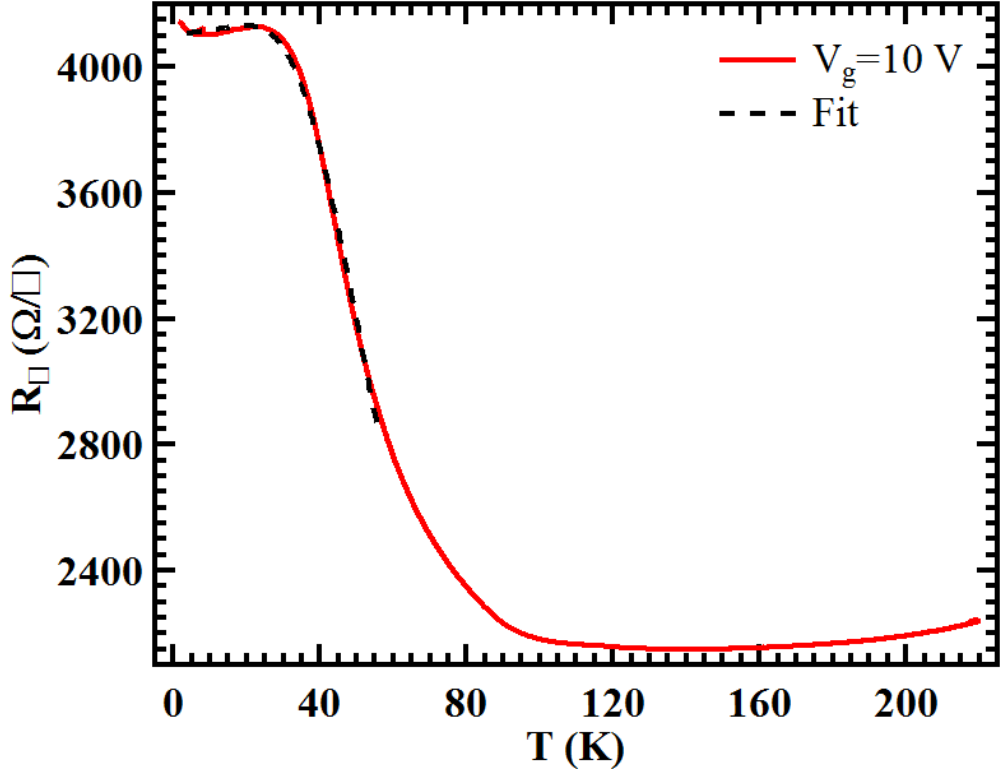

**Figure S1** Sheet resistance at zero *B*-field measured as a function of temperature (*T*) with $V_g$=10 V (duplicated from Fig. 1c). We have fitted our experimental data (red solid line) below 55 K to a simple model (back dashed line is the fitting result) similar to that developed in Ref. S1 to estimate the surface to bulk conductance ratio. This model assumes the total conductance $G_t$ consists of two parallel channels: a thermally activated bulk with conductance $G_b$, and a metallic conduction ($G_S$) which we attribute to the surface. At low T, the bulk conduction freezes out, and the residual metallic surface conduction dominates. More specifically, we fit the measured $G_t(T)$ to $G_t(T) = G_b(T) + G_s(T)$, with $G_b(T) = 1/(R_{bo}e^{\Delta/T})$ and $G_S(T) = 1/(A + BT)$, where Δ represents a bulk activation gap between the Fermi level and the edge of the bulk band, *A* accounts for the static disorder scattering and *B* represents electron-phonon scattering affecting the surface carriers (Ref. S1). The fit of the data yields Δ=200 K, $R_{b0}$=246.3 Ω/□, *A*=4100 Ω/□, and *B*=1.68 Ω/(□K). Based on these values, one obtains $R_b$(*T*=20 K)=193 kΩ/□ and $R_S$(*T*=20 K)=4.1 kΩ/□, giving a surface to bulk conductance ratio of $G_S/G_b=R_b/R_S$~50 at *T*=20 K, suggesting the surface conduction will dominate the charge transport at low temperature ($G_S/G_b$>50 for *T*<20 K). The insulating behavior in *R* vs *T* fitted above is dominated by the bulk carrier freeze-out as captured by this model, and the effect of *T*-dependence of the STO gate capacitance (not considered in this simple model) on the resistance in the fitted *T* range is relatively small (as can be seen by comparing the $V_g$=10 V data and $V_g$=130 V data in Fig. 1c) and will not significantly change the fitting results or affect our conclusion.

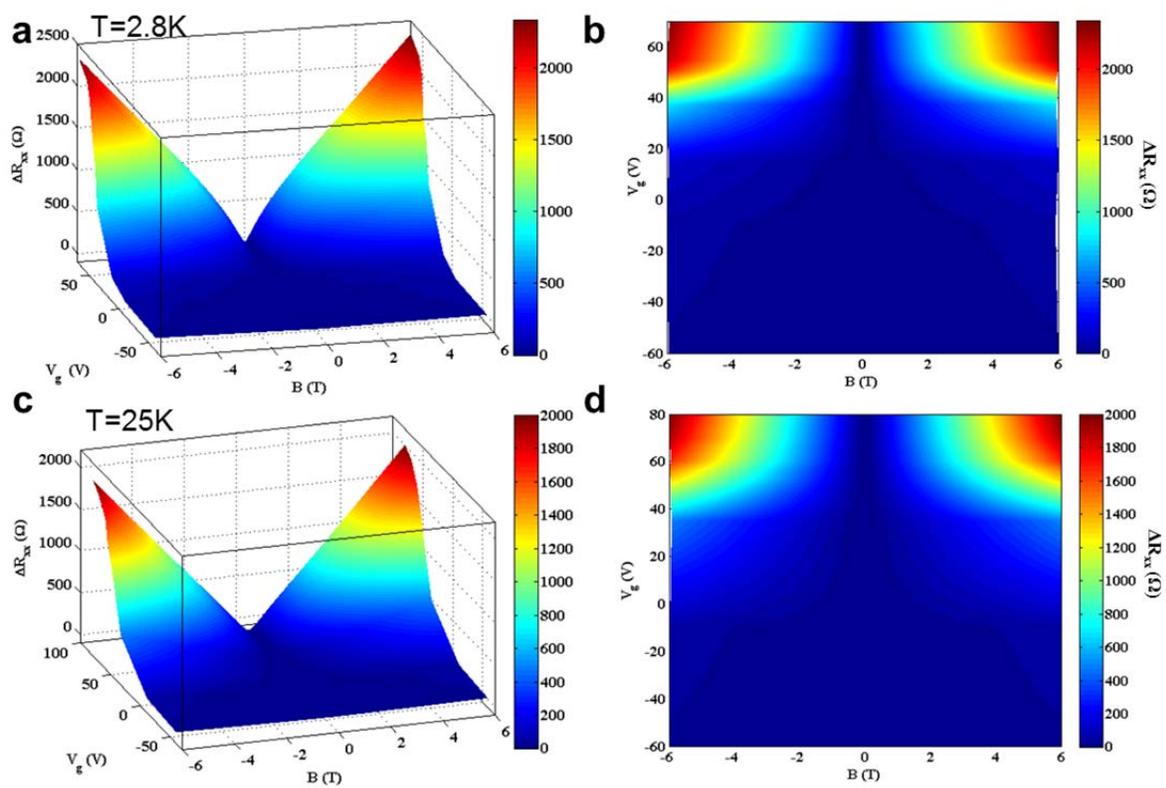

**Figure S2** Color plots of the magnetoresistance $\Delta R_{xx}=R_{xx}(B)-R_{xx}(B=0T)$ vs $B$ and $V_g$ at two different temperatures: (a) 3D color surface plot and (b) 2D color plot for $T$=2.8 K; (c) 3D color surface plot and (d) 2D color plot for $T$=25 K.

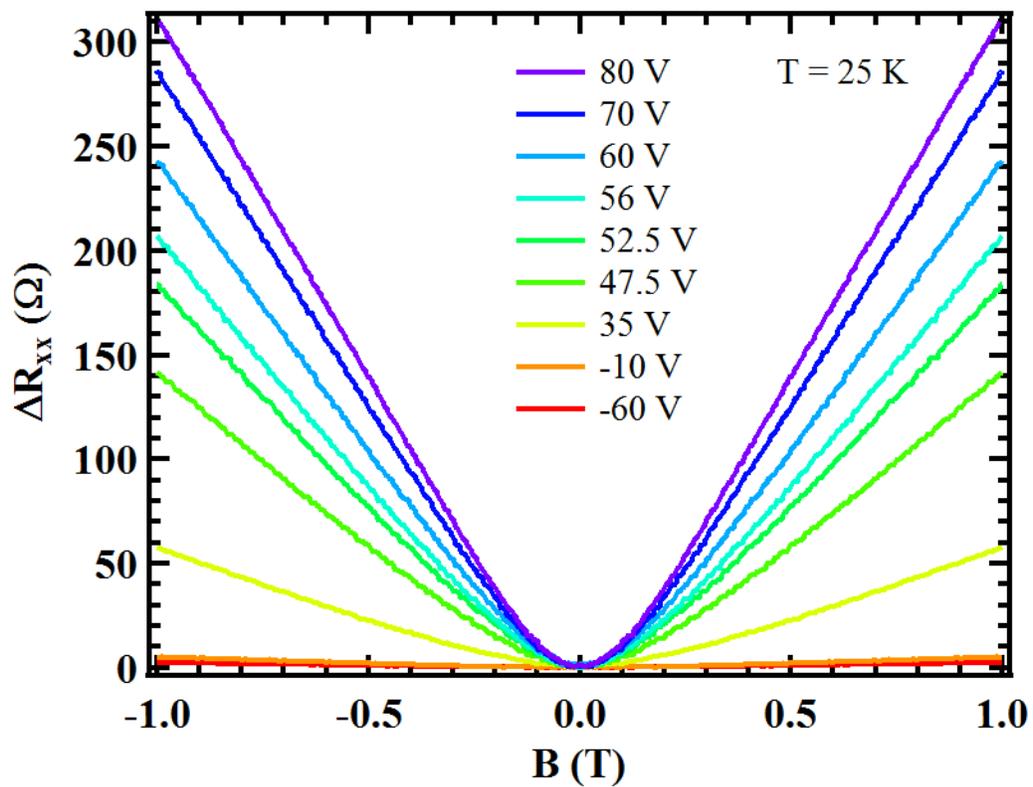

**Figure S3** A zoomed in view of the upper panel of Fig. 2c showing the magnetoresistance $\Delta R_{xx}=R_{xx}(B)-R_{xx}(B=0T)$ vs $B$ for different $V_g$ at $T=25$ K, where LMR becomes prominent and starts from very low $B$ (<~ 0.2 T) for most of $V_g$'s.

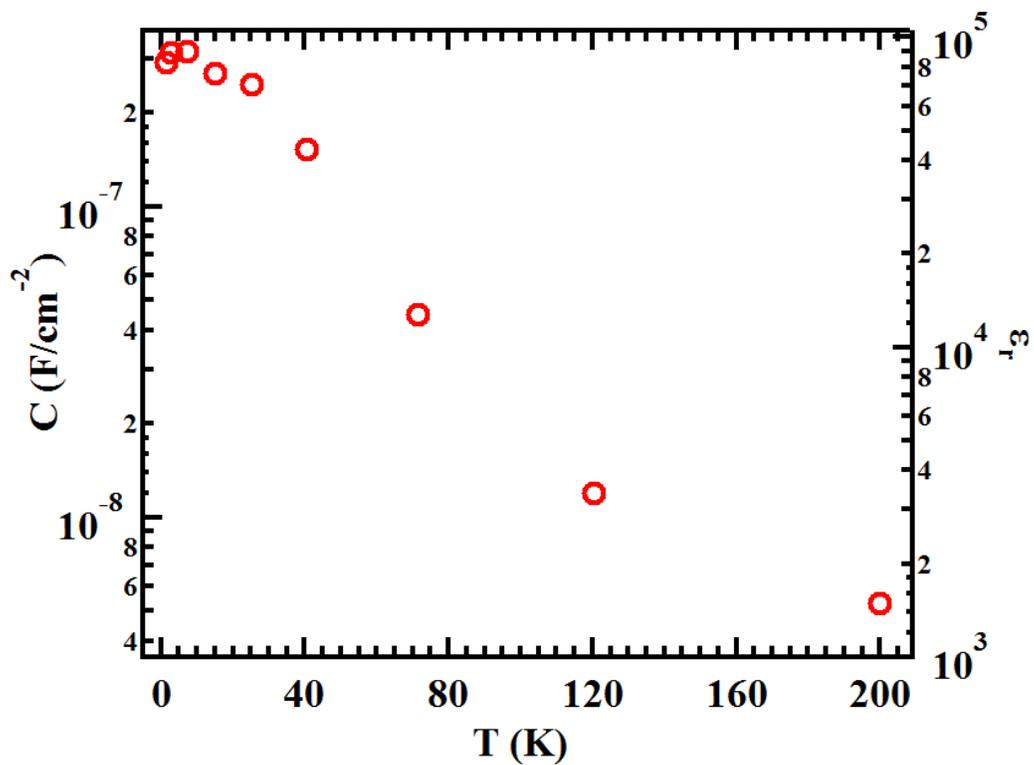

**Figure S4** The extracted gate capacitance $C$ per unit area (from Fig. 3c) and effective relative dielectric constant $\varepsilon_r$ of the 250 μm-thick STO substrate vs $T$.

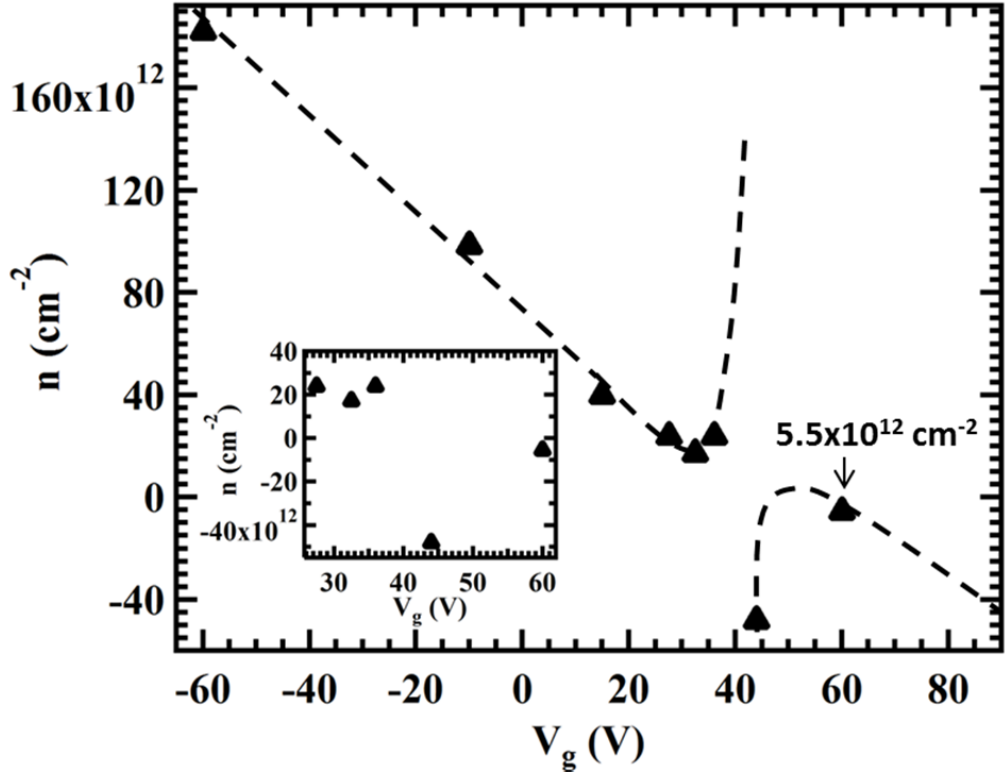

**Figure S5** The gate dependence of the carrier density $n$ extracted using a 1-band model from Fig. 2a at $T$=1.4 K. For $V_g$ far away from CNP (~40 V), $R_{xy}(B)$ is linear (Fig. 2a) and $n$ has a linear dependence on $V_g$. As $V_g$ is close to CNP for both n-type and p-type regimes, $R_{xy}(B)$ becomes nonlinear (Fig. 2a) due to multiple-band transport (eg., coexistence of electron and hole carriers), the $n$ (extracted from 1-band model using the low-$B$ slope of $R_{xy}$) deviates from the linear $V_g$ dependence and its magnitude provides an *upper bound* for the actual net carrier density [S2]. The dashed line indicates the expected general trend of one-band-extracted $n$ vs $V_g$ as the system goes through the ambipolar transition [S2]. Inset is the zoom-in of $n$ vs $V_g$ near the CNP. We estimate the maximum surface carrier (hole) density achieved when the Fermi level ($E_f$) is at the top of BVB to be $n$~$1.2\times10^{13}$ cm$^{-2}$, for two surfaces combined), using $n=k_F^2/2\pi$, where $k_F$ is the Fermi wavevector based on the ARPES measured band structure (near BVB) in Ref. S3. It is expected that the Dirac point of the TSS is closer to the BVB than to the bulk conduction band (BCB, Ref. S3, even though the detailed band structure near the bottom of BCB has not been measured), suggesting that the maximum carrier (electron) density that can be accommodated by two surfaces (when $E_f$ is at the bottom of BCB) is even larger than ~$1.2\times10^{13}$ cm$^{-2}$.

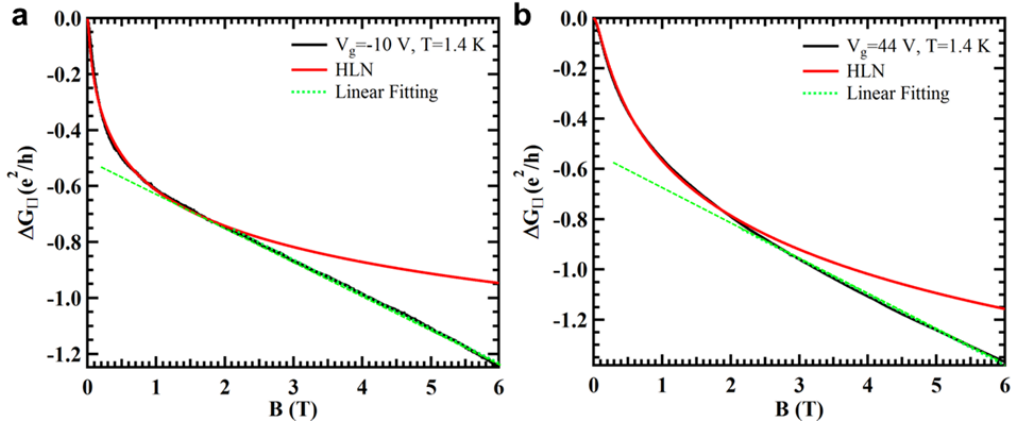

**Figure S6** Representative sheet conductance ($\Delta G_□ = G_□(B) - G_□(B=0T)$) as a function of $B$ field, for (a) $V_g$=-10 V and (b) $V_g$=44 V, both measured at $T$=1.4 K. Each red curve is the fit using the HLN model (Eq. 1) in the low-$B$ regime (up to 2 T) and green dotted line is the linear fitting of the LMR at high-$B$ regime (from 2 T to 6 T).

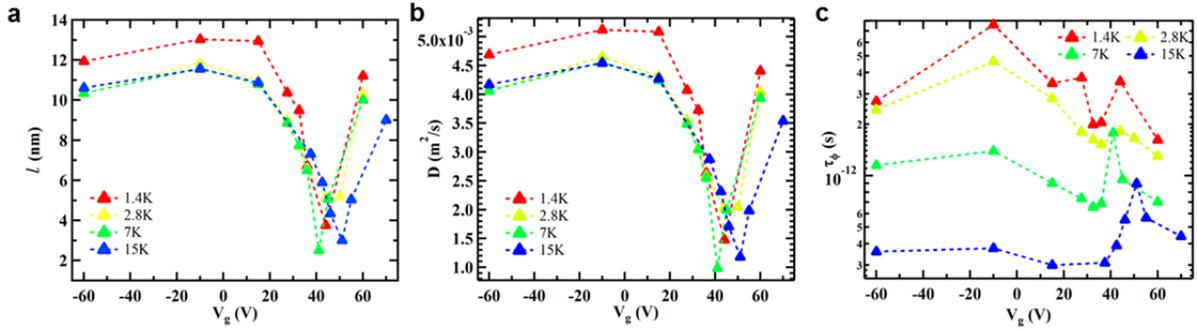

**Figure S7** Estimated gate-dependent (a) mean free path $l$, (b) diffusion constant $D$, and phase coherence time $\tau_\phi$ for surface state carriers, at different temperatures. The mean free path $l$ is estimated from $l=\sigma h/(e^2 k_F)$ with Fermi wave vector $k_F = \sqrt{2\pi n}$ using measured zero-$B$-field 2D electrical conductivity $\sigma$ (=$G_□$) and 1-band total Hall density $n$; diffusion constant $D$ from $D=v_F \times l$ with $v_F=4\times10^5$ m/s (based on ARPES measurement value [S3]), and phase coherence time $\tau_\phi$ using $\tau_\phi=L_\phi^2/D$ (with $L_\phi$ from Fig. 4c). We note that for $V_g$<0V, there will be significant carriers from BVB. However, we find that recalculated D and $\tau_\phi$ using adjusted $v_F$ (estimated from BVB parameters) are still on the same order of magnitude (within factor of 2) from the above estimates considering only the surface state carriers, and do not change our conclusions.

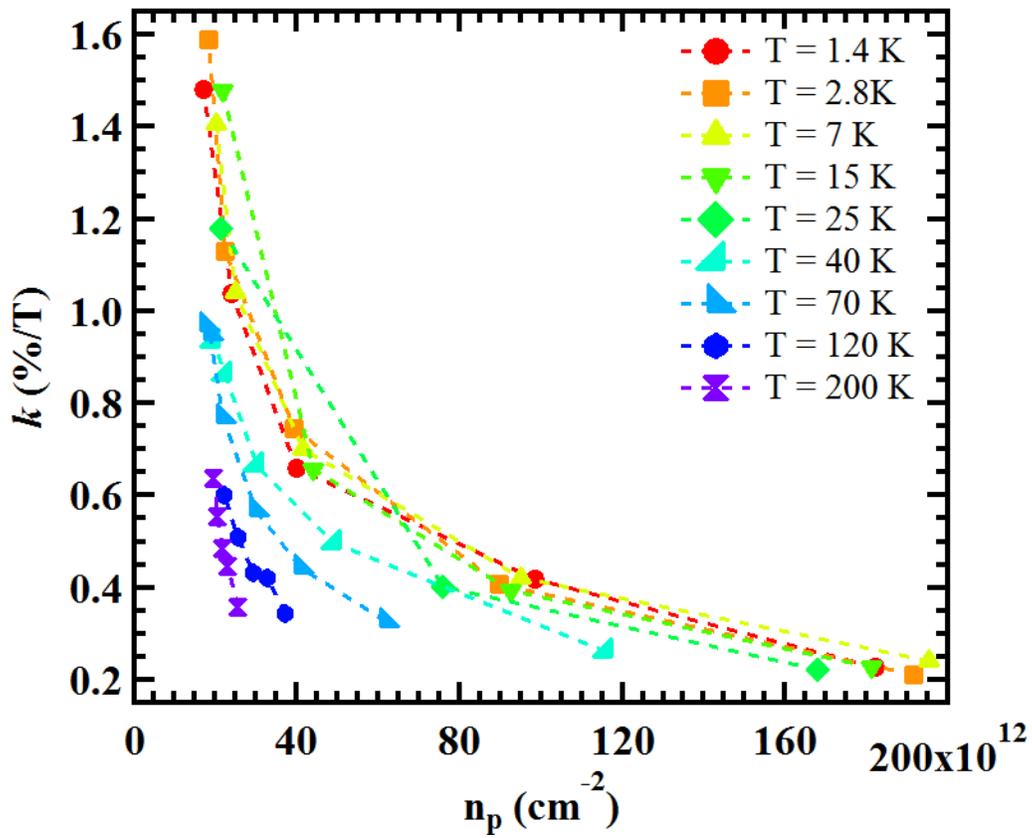

**Figure S8** The slope $k$ of LMR vs $n_p$ (of one-band p-type carriers mostly from BVB) at different temperatures, demonstrating the LMR is significantly enhanced as $E_f$ approaches TSS.

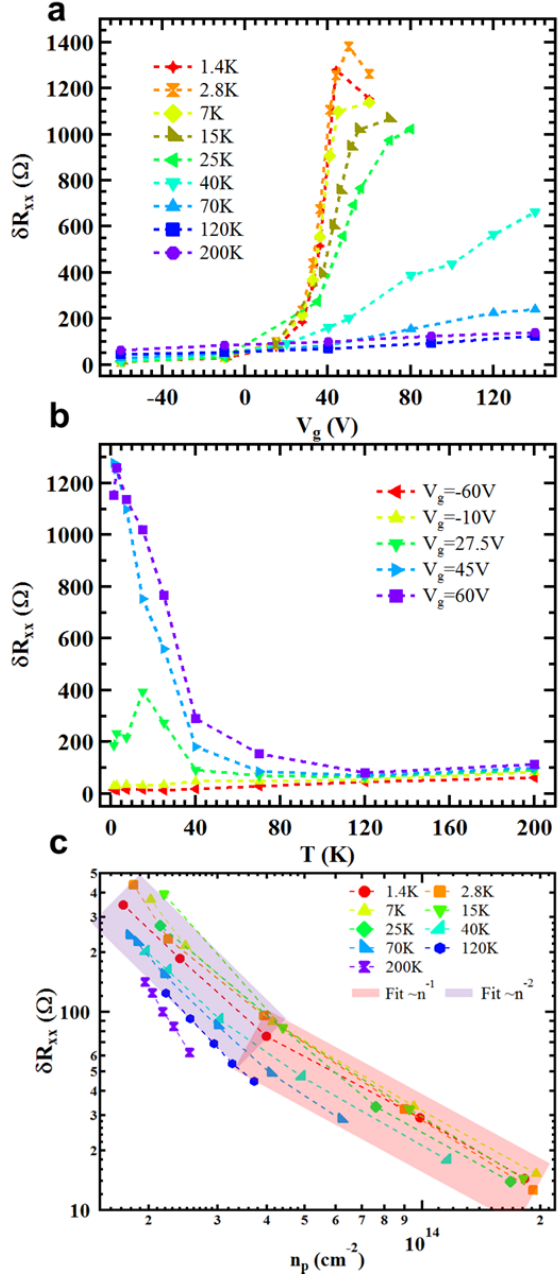

**Figure S9** (a) The $\delta R_{xx} = R_{xx}(6T) - R_{xx}(3T)$ vs $V_g$ at different temperatures, demonstrating the LMR is significantly enhanced as $E_f$ approaches TSS; (b) $\delta R_{xx}$ vs $T$ at five representative gate voltages; (c) $\delta R_{xx}$ vs $n_p$ (for one-band p-type carriers mostly from BVB, same $n_p$ data points as in Fig. 3c) at different temperatures. We find that $\delta R_{xx}$ cannot be fitted to a single power law (either $1/n_p^2$ (Abrikosov) or $1/n_p$ (Wang-Lei)) over this density range, but rather appear to cross over from a $\sim 1/n_p$ behavior (indicated by the pink band) for $n_p > 4 \times 10^{13}$ cm$^{-2}$ to $\sim 1/n_p^2$ behavior (purple band) for $n_p < 4 \times 10^{13}$ cm$^{-2}$ (except for the highest-$T$ data at 200K).

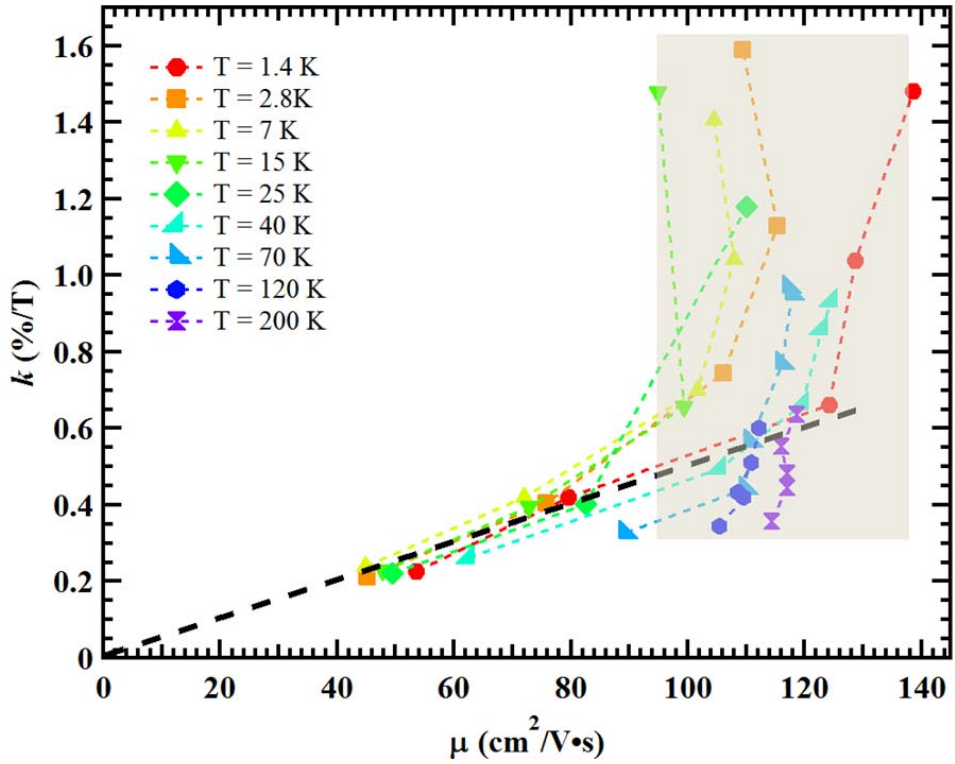

**Figure S10** The slope $k$ of LMR vs mobility $\mu$ (for one-band p-type carriers mostly from BVB) at different temperatures. When $\mu$ is smaller than ~100 cm$^2$/Vs, $k$ is approximately proportional to $\mu$, while $k$ rise sharply and becomes less well correlated with $\mu$ when $\mu$ approaches ~100-130 cm$^2$/Vs (highlighted by the grey band). We note this "threshold" mobility is close to the "maximum" $\mu_0$ fitted in the $\mu$ vs. $n_p$ dependence (inset of Fig. 3d).

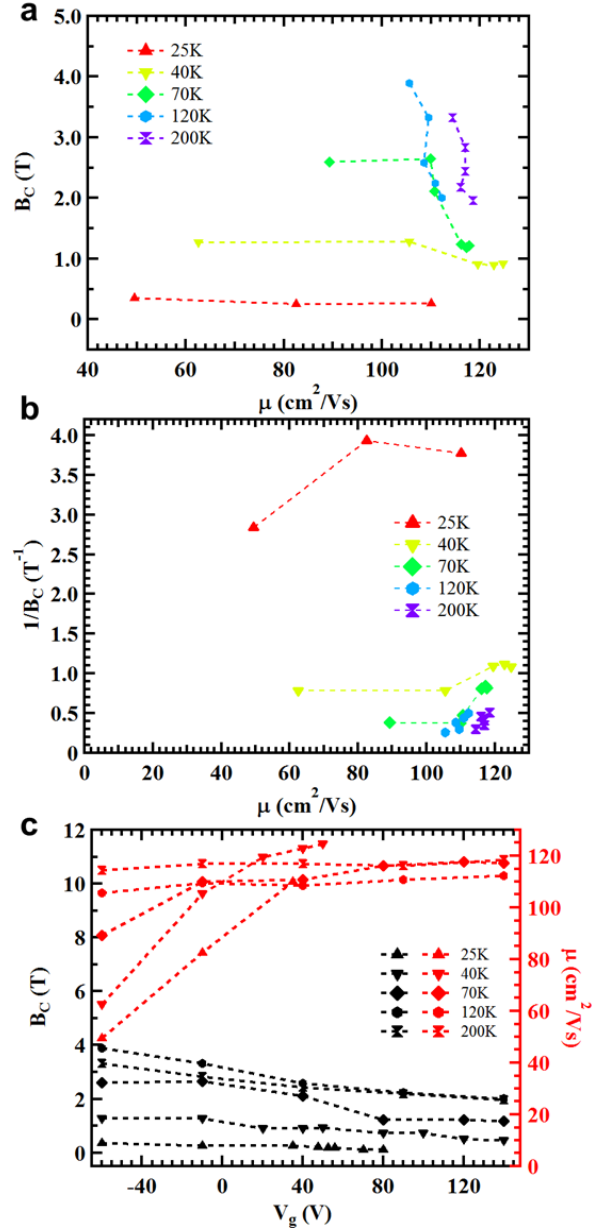

**Figure S11** (a) The crossover field $B_C$ as a function of carrier mobility $\mu$; (b) $1/B_C$ as functions of $\mu$; (c) the crossover field $B_C$ on left axis and Hall mobility $\mu$ on right axis as a function of $V_g$ at various temperatures. The crossover field $B_C$ (obtained from the first-order derivative of the MR curve versus $B$ field) is the field at which the MR curve changes from parabolic to linear (i.e., MR derivative approaching a constant). The data points involve $\mu$ only include those for one-band p-type carriers.

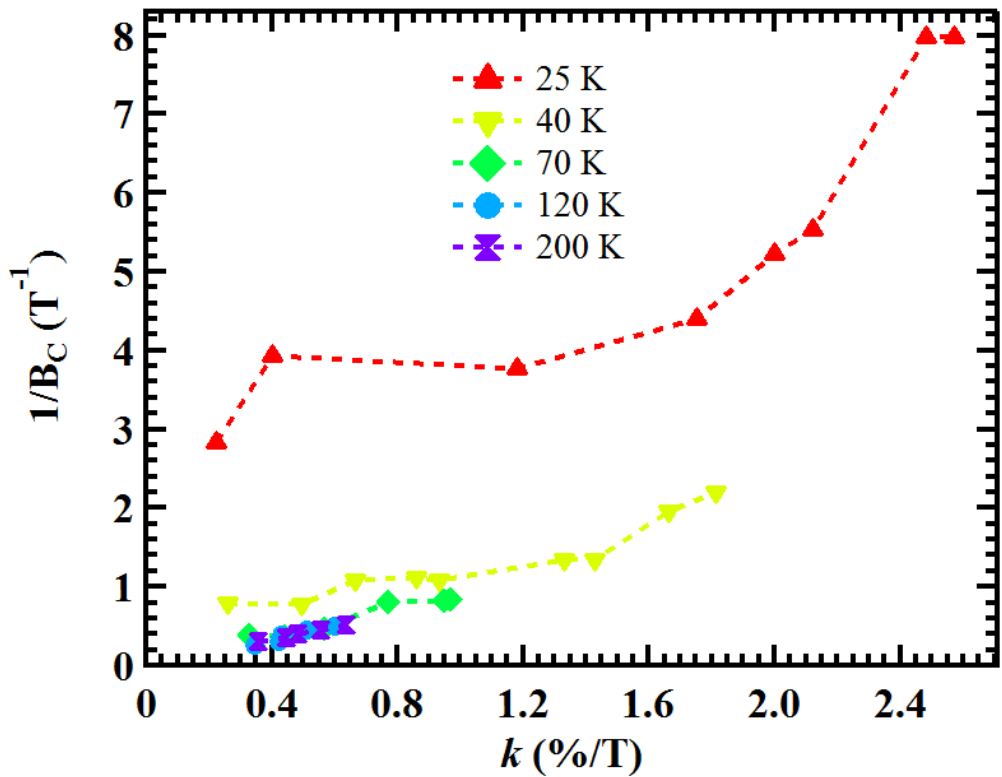

**Figure S12** The $1/B_C$ as a function of the slope $k$ of LMR at various temperatures for all gate voltage values measured.